\documentclass[a4paper,10pt,twoside]{cpc-hepnp}

\usepackage{multicol}
\usepackage{multirow}
\usepackage{graphicx}
\usepackage{booktabs}
\usepackage{amssymb,bm,mathrsfs,bbm,amscd}
\usepackage[tbtags]{amsmath}
\usepackage{lastpage}

\begin{document}

\fancyhead[c]{\small Chinese Physics C~~~Vol. XX, No. X (201X)
XXXXXX} \fancyfoot[C]{\small 010201-\thepage}

\footnotetext[0]{Submitted 25 March 2016}

\title{Searching for high-$K$ isomers in the proton-rich $A\sim80$ mass region
\thanks{
Supported by National Key Basic Research Program of China (2013CB834402) and National Natural Science Foundation of China (11235001, 11320101004 and 11575007)}}

\author{
      Zhi-Jun Bai$^{1}$
\quad Chang-Feng Jiao$^{2}$
\quad Yuan Gao$^{1}$
\quad Fu-Rong Xu$^{1;1)}$\email{frxu@pku.edu.cn}
}
\maketitle

\address{%
$^1$ State Key Laboratory of Nuclear Physics and Technology, School of Physics, Peking University, Beijing 100871, China\\
$^2$ Department of Physics and Astronomy, University of North Carolina, Chapel Hill, North Carolina, 27599-3255, USA\\
}

\begin{abstract}
  Configuration-constrained potential-energy-surface calculations have been performed to investigate the $K$ isomerism in the proton-rich $A\sim80$ mass region. An abundance of high-$K$ states are predicted. These high-$K$ states arise from two and four-quasi-particle excitations, with $K^{\pi}=8^{+}$ and $K^{\pi}=16^{+}$, respectively. Their excitation energies are comparatively low, making them good candidates for long-lived isomers. Since most nuclei under studies are prolate spheroids in their ground states, the oblate shapes of the predicted high-$K$ states may indicate a combination of $K$ isomerism and shape isomerism.
\end{abstract}



\footnotetext[0]{\hspace*{-3mm}\raisebox{0.3ex}{$\scriptstyle\copyright$}2013
Chinese Physical Society and the Institute of High Energy Physics
of the Chinese Academy of Sciences and the Institute
of Modern Physics of the Chinese Academy of Sciences and IOP Publishing Ltd}%

\begin{multicols}{2}

\section{Introduction}
Isomers, as metastable states of atomic nuclei which are arguably the most complex quantum many-body systems, have provided valuable insights into nuclear structure and played a unique role in the development of nuclear structural theories. Recently, they have also been shown to be instrumental in the advancement to the island of stability \cite{HGB06}. Practically, efforts are well under way to harness their potential as a form of energy storage \cite{WD99}.

In terms of underlying mechanisms, there are three types of nuclear isomers: shape isomers \cite{AR09,WX06}, spin traps \cite{XW00,DL06} and $K$ traps \cite{JD09,XZ04,LX11}. The first type arises when a nucleus can have distinct shapes and the difference between those shapes is large enough to effectively block the transition from one shape to another. The two latter types owe their occurrence to selection rules of angular momentum in electromagnetic transition. Specifically, transitional forbiddance due to a substantial change in the {\it magnitude} of nuclear spin leads to spin traps, while that due to a substantial change in the {\it orientation} of nuclear spin leads to $K$ traps \cite{WX16}.

The proton-rich $A\sim80$ mass region has been under intensive studies for its intricate manifestation of shape coexistence \cite{WHN92,BFJ15} and importance in the $rp$ process \cite{SAG98,SAB01,KFF01}. Incidentally, $^{72}$Kr turns out to be another good example of shape isomerism \cite{BMK03,MSB09}, as distinguished from fission isomers. Less discussed is the $K$ isomerism in this region \cite{SUN04}. In fact, the relevant Nilsson diagram shows that this mass region is rich in large subshell gaps \cite{BFJ15}. These subshell gaps not only play a decisive role in the stabilization of deformation by providing extra energy reduction in shell correction, but also facilitate the emergence of quasi-particle states through a similar mechanism. The present work searches this mass region for candidates for high-$K$ isomers by means of configuration-constrained potential-energy-surface (PES) calculations, which have proved to be a successful approach to the description of quasi-particle states \cite{WX16,XWS98}.

\section{Theoretical framework}
The configuration-constrained potential-energy surface method \cite{XWS98} is a macroscopic-microscopic approach that is aimed at the description of the multi-quasi-particle excitation of nuclei. It employs the standard liquid drop model \cite{MS66} as the macroscopic part and the Strutinsky shell correction \cite{St67} as the microscopic part. The latter, in detail, consists of single-particle energies, pairing interaction and Pauli-blocking effects.

Single-particle levels are obtained with a triaxially deformed Woods-Saxon potential \cite{NDB85,CDN87}, using the universal parameter set \cite{DSW81}. To properly account for nuclear superfluidity, monopole pairing is incorporated, whose strength $G$ is determined by the average gap method \cite{MN92,XW99}. An approximate particle-number projection, known as the Lipkin-Nogami method \cite{PNL73,NRG90,XWS98,XW99} is implemented in order to reduce the fluctuation in particle number and therefore avoid 
spurious pairing collapse.

At each grid point $(\beta_{2},\gamma)$ on the quadrupole deformation plane, the Lipkin-Nogami equation is solved and shell correction calculated, which, combined with the macroscopic energy gain due to deformation, gives the total potential energy of the nucleus. Then, this potential energy is further minimized with respect to the hexadecapole deformation 
$\beta_{4}$ to take into account possible high-order deformation. The minima on the obtained potential-energy surface correspond to equilibrium shapes of nuclei. So the potential-energy-surface method is pairing-deformation self-consistent. Moreover, by means of the averaged Nilsson quantum numbers, an identifying method has been devised \cite{XWS98} and makes it possible that certain single-particle orbitals can be traced and blocked for all calculated deformations, resulting in configuration-constrained potential-energy surfaces.

\section{Results and discussion}

\begin{center}
\includegraphics[width=8.cm]{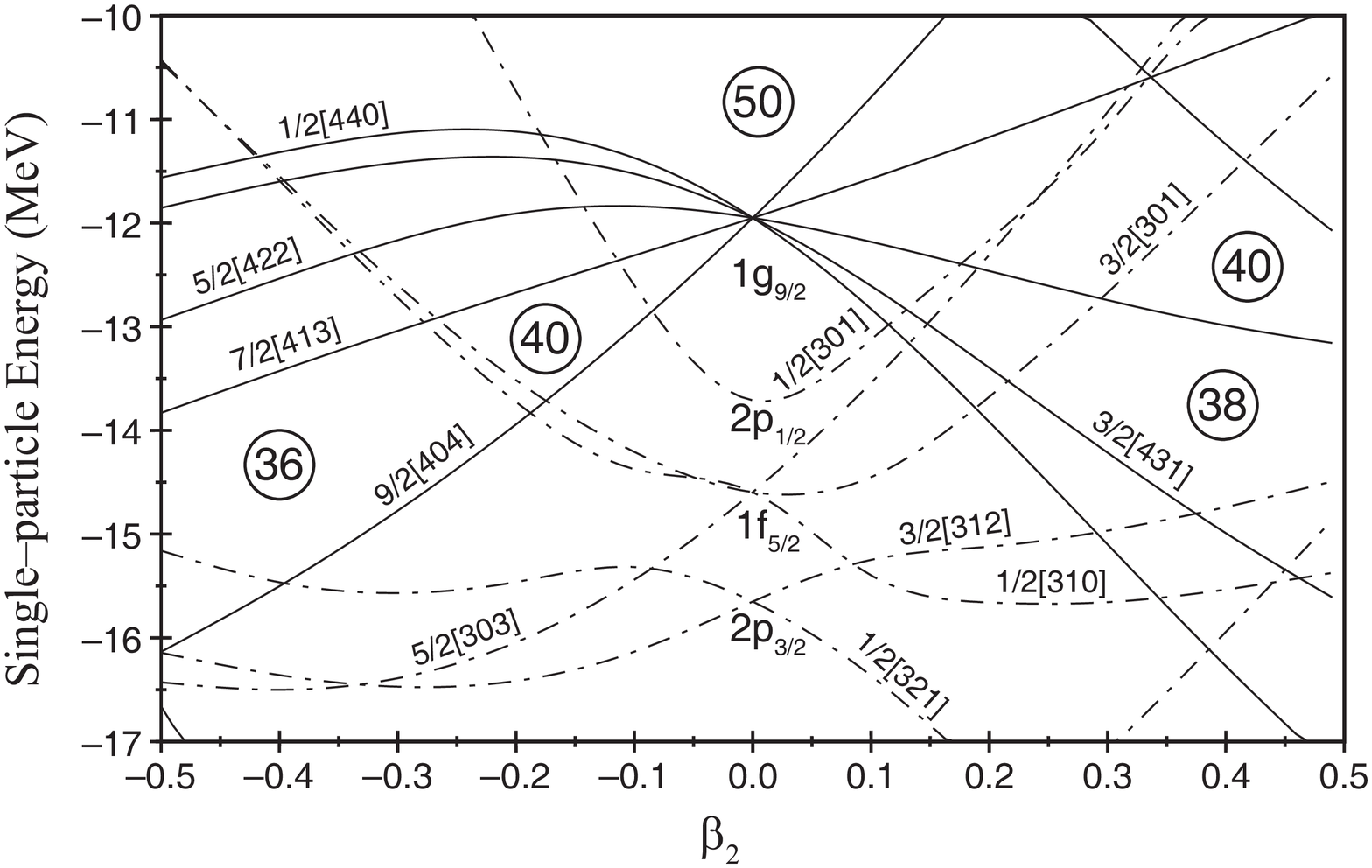}
\figcaption{\label{74Kr-Nil}The Nilsson diagram for the neutrons of $^{74}$Kr, derived from the Woods-Saxon potential \cite{NDB85,CDN87} with the universal parameter set \cite{DSW81}.
}
\end{center}

Single-particle levels of the neutrons of $^{74}$Kr are calculated with the Woods-Saxon potential and plotted in Fig.~\ref{74Kr-Nil}. The single-particle level scheme for its protons are similar, except that the levels are approximately 8 MeV higher because of coulomb force. Indeed, the Nilsson diagram in Fig.~\ref{74Kr-Nil} is typical for the $A\approx 80$ mass region. The abundance of subshell gaps complicates the scenarios of shape coexistence \cite{BFJ15} and yet favors the forming of quasi-particle states. In particular, in the splitting of the $g_{9/2}$ spherical subshell, two high-$\Omega$ orbitals, $\frac{7}{2}[413]$ and $\frac{9}{2}[404]$, dive rapidly with increased oblate deformation, giving rise to subshell gaps at $Z(N)=36,40$. It has been observed experimentally that the $g_{9/2}$ orbitals can play a key role in driving nuclei to different shapes \cite{DF07}.

Shown in Fig.~\ref{Kr08816} is the potential-energy surfaces for the ground and quasi-particle states of $^{72}$Kr. Owing to the proton and neutron subshell gaps at 36, the ground state of $^{72}$Kr is oblate deformed (see panel (a) of Fig.~\ref{Kr08816}), which has been corroborated by various experiments (see Ref.~\cite{ILM14} and references therein). Likewise, its two-quasi-proton ($K^{\pi}=8^{+}$) state with the configuration $\pi\{\frac{7}{2}^{+}[413],\frac{9}{2}^{+}[404]\}$, two-quasi-neutron ($K^{\pi}=8^{+}$) state with a similar configuration $\nu\{\frac{7}{2}^{+}[413],\frac{9}{2}^{+}[404]\}$, and four-quasi-particle ($K^{\pi}=16^{+}$) state with the configuration $\pi\{\frac{7}{2}^{+}[413],\frac{9}{2}^{+}[404]\}\otimes\nu\{\frac{7}{2}^{+}[413],\frac{9}{2}^{+}[404]\}$ are also well-deformed oblate spheroids (see the last three panels of Fig.~\ref{Kr08816}).

An analogous case occurs in $^{80}$Zr, except that the nucleus is here, in perfect agreement with experiment \cite{LCC87}, prolate deformed in the ground state (see Fig.~\ref{Zr08816}), due to a more significant subshell gap at $Z(N)=40$ in the prolate direction (see Fig.~\ref{74Kr-Nil}). The high-$K$ states persist although less deformed, with an oblate deformation of $\beta_{2}\sim-0.18$. This indicates that the unpaired nucleons could exert a strong shape-polarizing effect on the whole nucleus, driving it from a prolate shape in the ground state to an oblate shape in excitation. This effect has been discovered in other mass regions \cite{XWW99,SX10} and is recognized as a characteristic of quasi-particle excitation. In consequence, the potential-energy surfaces for the four-quasi-particle states look more rigid than those for the two-quasi-particle states (obvious in Fig.~\ref{Zr08816} but less so in Fig.~\ref{Kr08816}), signifying enhanced stability of deformation.
\end{multicols}

\begin{center}
\includegraphics[width=16cm]{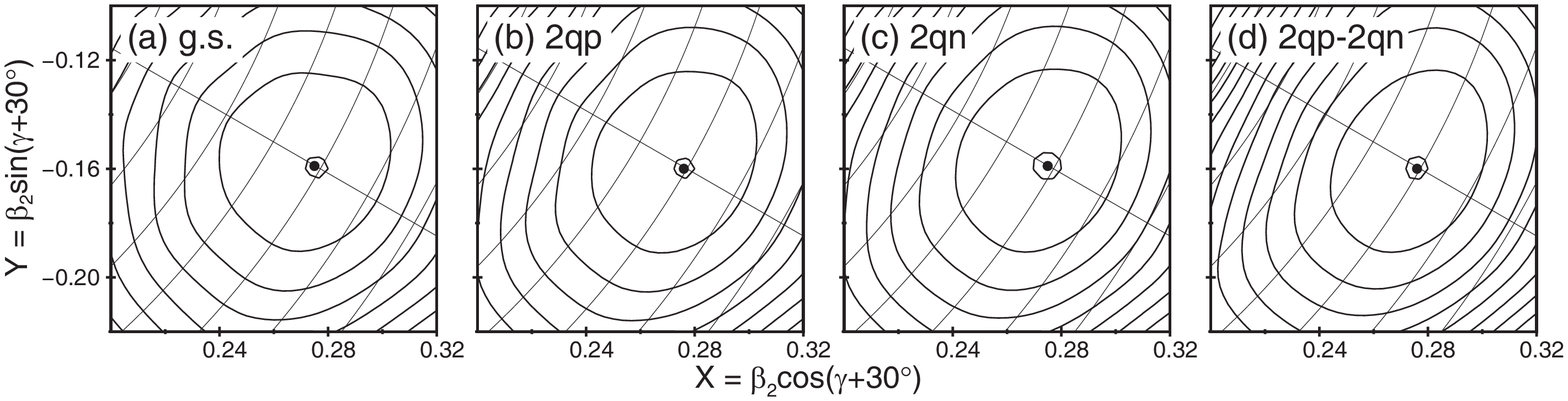}
\figcaption{\label{Kr08816}Potential-energy surfaces of $^{72}$Kr for its (a)~ground state (g.s.), (b)~two-quasi-proton state (2qp), (c)~two-quasi-neutron state (2qn) and (d)~two-quasi-proton-two-quasi-neutron state (2qp-2qn). The configurations of the excited states are $\pi\{\frac{7}{2}^{+}[413],\frac{9}{2}^{+}[404]\}$, $\nu\{\frac{7}{2}^{+}[413],\frac{9}{2}^{+}[404]\}$ and $\pi\{\frac{7}{2}^{+}[413],\frac{9}{2}^{+}[404]\}\otimes\nu\{\frac{7}{2}^{+}[413],\frac{9}{2}^{+}[404]\}$, respectively. Neighboring contours are 100~keV apart in energy.  The deformations and excitation energies of the minima can be found in Table~\ref{tab08816}.}
\end{center}

\begin{center}
\includegraphics[width=16.8cm]{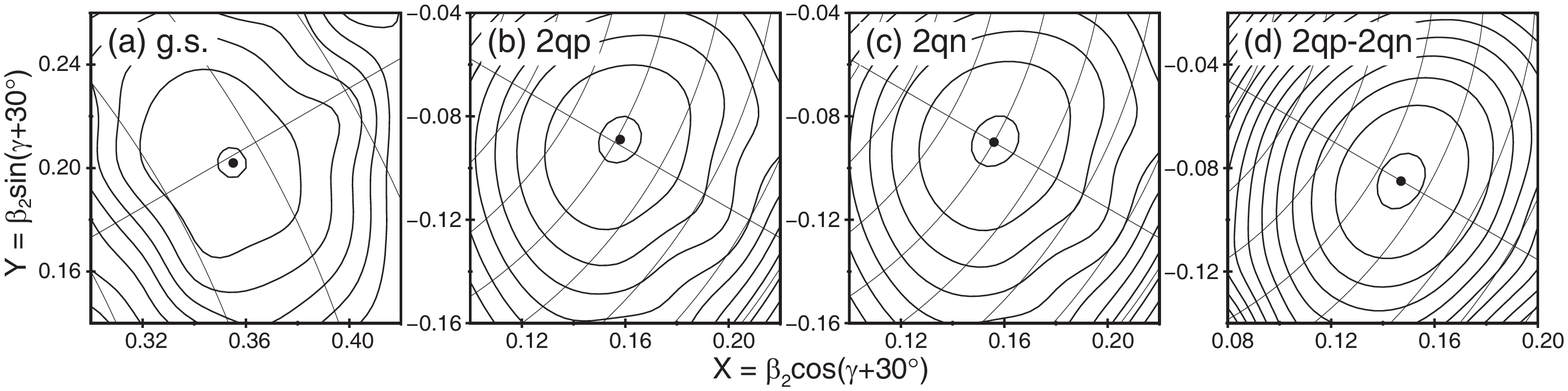}
\figcaption{\label{Zr08816}Similar to Fig.~\ref{Kr08816} but of $^{80}$Zr.}
\end{center}

\begin{table}[htb!]
  \centering
  \caption{Calculated ground, two-quasi-proton ($K^{\pi}=8^{+}$), two-quasi-neutron ($K^{\pi}=8^{+}$) and two-quasi-proton-two-quasi-neutron ($K^{\pi}=16^{+}$) states of proton-rich even-even isotopes in the $A\sim80$ mass region. The corresponding configurations of the excited states are $\pi\{\frac{7}{2}^{+}[413],\frac{9}{2}^{+}[404]\}$, $\nu\{\frac{7}{2}^{+}[413],\frac{9}{2}^{+}[404]\}$ and $\pi\{\frac{7}{2}^{+}[413],\frac{9}{2}^{+}[404]\}\otimes\nu\{\frac{7}{2}^{+}[413],\frac{9}{2}^{+}[404]\}$, respectively. $|\gamma|\approx0^{\circ}$ or $60^{\circ}$ in call cases, except for $^{78}$Mo whose ground state is rather $\gamma$-soft. Negative $\beta_{2}$ values denote oblate deformation. Otherwise, nuclei are prolate. Excitation energies are in MeV.}\label{tab08816}
  \addtolength{\tabcolsep}{-1.0 pt}
  \begin{tabular}{cccccccccccccccccc}
    \hline\hline
     \multirow{2}{*}{Isotope} & \multicolumn{2}{c}{ground state} && \multicolumn{3}{c}{two-quasi-proton} && \multicolumn{3}{c}{two-quasi-neutron} &&
     \multicolumn{3}{c}{four-quasi-particle} \\ \cline{2-3}\cline{5-7}\cline{9-11}\cline{13-15}
     & $\beta_{2}$ & $\beta_{4}$ && $\beta_{2}$ & $\beta_{4}$ & $E_{\mathrm{x}}$ && $\beta_{2}$ & $\beta_{4}$ & $E_{\mathrm{x}}$ && $\beta_{2}$ & $\beta_{4}$ & $E_{\mathrm{x}}$ \\
     \hline
     ${}^{70}$Se & $-$0.283 & \hphantom{$-$}0.001 && --- & --- & --- && $-$0.289 & $-$0.005 & 2.506 && --- & --- & --- \\
     ${}^{72}$Se & $-$0.269 & $-$0.003 && --- & --- & --- && $-$0.248 & $-$0.016 & 2.525 && --- & --- & --- \\
     ${}^{70}$Kr & $-$0.284 &\hphantom{$-$}0.002 && $-$0.290 & $-$0.004 & 2.475 && --- & --- & --- && --- & --- & --- \\
     ${}^{72}$Kr & $-$0.313 &\hphantom{$-$}0.004 && $-$0.319 & $-$0.003 & 2.452 && $-$0.318 & $-$0.003 & 2.484 && $-$0.318 & $-$0.010 & 4.921 \\
     ${}^{74}$Kr & \hphantom{$-$}0.375 & \hphantom{$-$}0.014 && $-$0.318 & $-$0.006 & 3.197 && $-$0.259 & $-$0.015 & 3.538 && $-$0.276 & $-$0.019 & 6.037 \\
     ${}^{76}$Kr & \hphantom{$-$}0.376 & \hphantom{$-$}0.002 && $-$0.285 & $-$0.016 & 3.231 && $-$0.190 & $-$0.033 & 3.297 && $-$0.239 & $-$0.031 & 6.117 \\
     ${}^{78}$Kr & $-$0.244 & $-$0.021 && $-$0.266 & $-$0.025 & 2.536 && --- & --- & --- && $-$0.233 & $-$0.029 & 6.082 \\
     ${}^{72}$Sr & \hphantom{$-$}0.381 & \hphantom{$-$}0.031 && $-$0.252 & $-$0.015 & 2.832 && --- & --- & --- && --- & --- & --- \\
     ${}^{74}$Sr & \hphantom{$-$}0.376 & \hphantom{$-$}0.016 && $-$0.266 & $-$0.015 & 3.647 && $-$0.319 & $-$0.006 & 3.301 && $-$0.280 & $-$0.020 & 6.119 \\
     ${}^{76}$Sr & \hphantom{$-$}0.388 & $-$0.001 && $-$0.253 & $-$0.022 & 4.544 && $-$0.250 & $-$0.021 & 4.583 && $-$0.239 & $-$0.034 & 6.955 \\
     ${}^{78}$Sr & \hphantom{$-$}0.394 & $-$0.012 && $-$0.235 & $-$0.034 & 4.263 && $-$0.181 & $-$0.039 & 4.242 && $-$0.199 & $-$0.043 & 6.646 \\
     ${}^{80}$Sr & \hphantom{$-$}0.378 & $-$0.010 && $-$0.232 & $-$0.037 & 2.839 && --- & --- & --- && $-$0.188 & $-$0.041 & 5.973 \\
     ${}^{76}$Zr & \hphantom{$-$}0.382 & \hphantom{$-$}0.003 && $-$0.230 & $-$0.031 & 3.869 && $-$0.300 & $-$0.014 & 3.561 && $-$0.242 & $-$0.031 & 6.501 \\
     ${}^{78}$Zr & \hphantom{$-$}0.396 & $-$0.012 && $-$0.187 & $-$0.038 & 4.605 && $-$0.236 & $-$0.030 & 4.582 && $-$0.202 & $-$0.043 & 6.980 \\
     ${}^{80}$Zr & \hphantom{$-$}0.405 & $-$0.022 && $-$0.181 & $-$0.044 & 4.110 && $-$0.180 & $-$0.043 & 4.140 && $-$0.173 & $-$0.052 & 6.187 \\
     ${}^{82}$Zr & \hphantom{$-$}0.397 & $-$0.021 && $-$0.183 & $-$0.048 & 2.479 && --- & --- & --- && $-$0.149 & $-$0.045 & 5.201 \\
     ${}^{78}$Mo & \hphantom{$-$}0.351 & \hphantom{$-$}0.002 && --- & --- & --- && $-$0.276 & $-$0.024 & 2.978 && $-$0.240 & $-$0.031 & 6.506 \\
     ${}^{80}$Mo & \hphantom{$-$}0.388 & $-$0.013 && --- & --- & --- && $-$0.235 & $-$0.036 & 3.559 && $-$0.195 & $-$0.042 & 6.690 \\
     ${}^{82}$Mo & \hphantom{$-$}0.412 & $-$0.023 && --- & --- & --- && $-$0.186 & $-$0.048 & 3.020 && $-$0.155 & $-$0.048 & 5.707 \\
    \hline\hline
  \end{tabular}
  \addtolength{\tabcolsep}{1.0 pt}
\end{table}

\begin{multicols}{2}

In fact, we have searched systematically the proton-rich side of the $A\sim80$ mass region for candidate nuclei, in which quasi-particle states with the configurations above could emerge. The results are tabulated in Table~\ref{tab08816}. Apparently, most calculated isotopes are prolate in their ground states. Yet through the shape-polarizing effect of unpaired nucleons, a wealth of high-$K$ states with oblate deformations are suggested by our calculations. The excitation energies of two-quasi-particle states ($K^{\pi}=8^{+}$) are generally around 3~MeV and those of four-quasi-particle states ($K^{\pi}=16^{+}$) around 6~MeV. Experimentally, the lowest-lying $J^{\pi}=8^{+},16^{+}$ states that have been observed in this mass region are typically at about 3~MeV and 8~MeV ($^{78}$Kr), respectively \cite{NNDC}. This means that the high-$K$ states listed in Table~\ref{tab08816}, if they do exist, have a good chance of being yrast. Even if they are not, the transition deexciting them are almost certainly $K$-forbidden, leading to high-$K$ isomers. Besides, for those nuclei which are prolate spheroids in their ground states, the deexcitation of the high-$K$ states are probably further hindered due to shape alteration.

\begin{center}
\includegraphics[width=8.5cm]{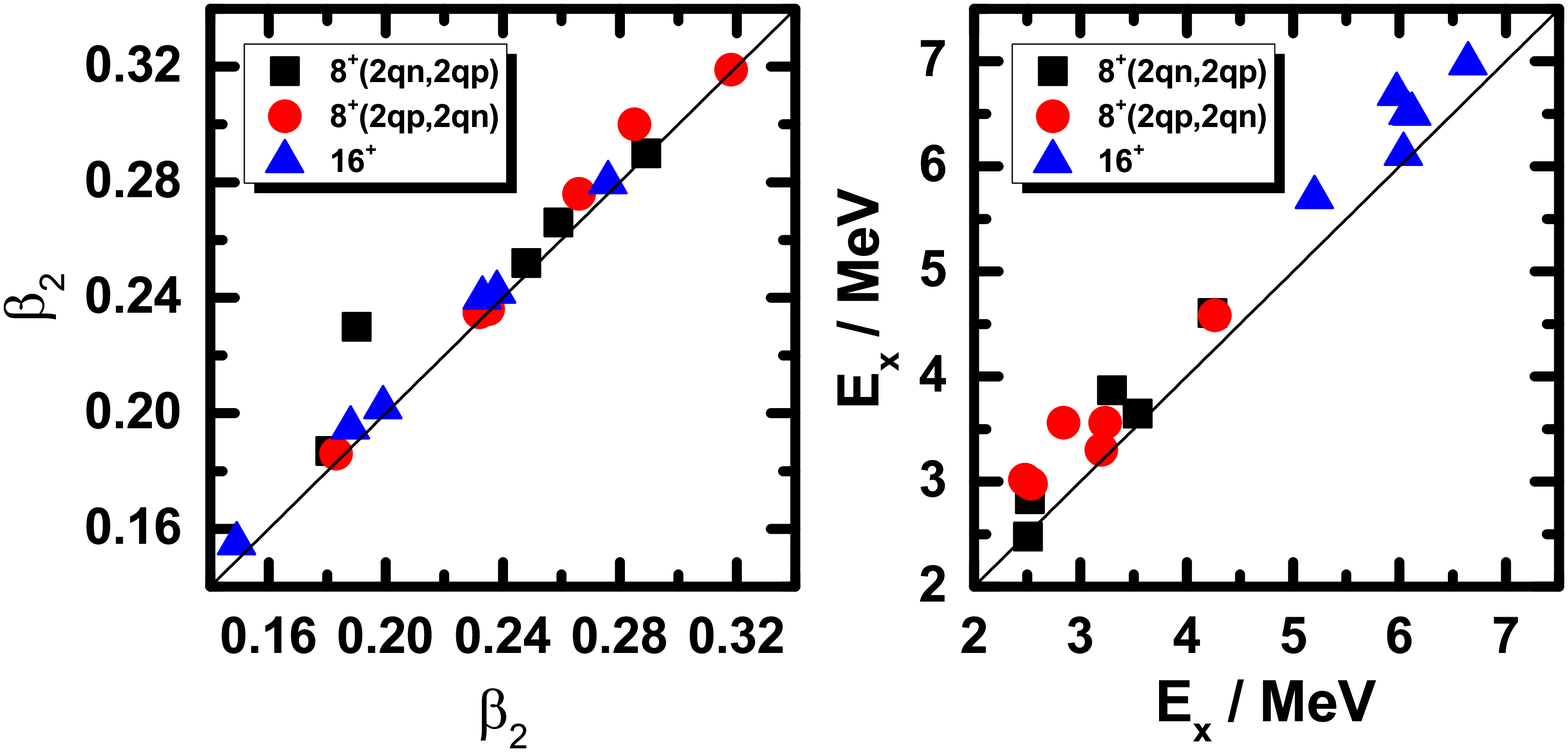}
\figcaption{\label{b2b2exex}(Color online) Comparison of deformations and excitation energies of the predicted high-$K$ states between mirror nuclei. The horizontal axes are the quantities of isotopes with $N>Z$, whereas the vertical ones are the quantities of their mirror partners. The $K^{\pi}=8^{+}$ states in Table~\ref{tab08816} are divided into two groups, with the two-quasi-neutron states of isotopes with $N>Z$ versus the two-quasi-proton states of their mirror partners in one group (labelled as `$8^{+}$(2qn,2qp)' in the legend) and the reverse in the other (labelled as `$8^{+}$(2qp,2qn)').}
\end{center}

An interesting observation about Table~\ref{tab08816} is the symmetry between mirror pairs, which is shown more clearly in Fig.~\ref{b2b2exex}. The two-quasi-neutron states of isotopes resemble the two-quasi-proton states of their mirror partners and vice versa, especially in terms of deformation. The same is also true of the $K^{\pi}=16^{+}$ states. This isospin symmetry could be traced to the universal parameter set \cite{DSW81} of the Woods-Saxon potential, which was fitted to experimental data across the nuclear chart. In this parameter set the values for protons and neutrons are very close.  The tiny but systematic upshift of the excitation-energy data points (see the right panel of Fig.~\ref{b2b2exex}) probably has something to do with Coulomb force.

There are many possible reasons why none of such isomers have been found yet. One relevant fact is that many of the calculated nuclides are rather proton-rich and some have even not been discovered ($^{76}$Zr, for instance), so accumulated experimental information on them is comparatively scant. Another reason is that experimental efforts directed to this mass region have been mainly focused on the phenomenon of shape coexistence \cite{BFJ15} and the roles that nuclei in this region play in astrophysical processes \cite{KFF01}. In addition, the predicted isomers are likely to be very long-lived and the gamma rays deexciting them therefore could be both delayed and subdued. So it cannot be ruled out that the predicted high-$K$ states might have eluded all detections so far.

\section{Summary}

$K$ isomerism in the proton-rich $A\sim80$ mass region has been studied by means of configuration-constrained potential-energy surfaces. Our calculations predict a large quantity of high-$K$ states. Two quasi-protons or quasi-neutrons with the configuration $\{\frac{7}{2}^{+}[413],\frac{9}{2}^{+}[404]\}$ lead to $K^{\pi}=8^{+}$ states, while with both protons and neutrons excited from vacuum, $K^{\pi}=16^{+}$ states occur. These high-$K$ states have comparatively low excitation energies, making them potential long-lived isomers. Since most of the calculated isotopes are prolate deformed in their ground states, the oblate deformation of the high-$K$ states suggests a combination of $K$ isomerism and shape isomerism.

\acknowledgments{Discussion with LIU H L, Xi'an Jiaotong University is gratefully acknowledged.}

\end{multicols}

\vspace{10mm}

\vspace{-1mm}
\centerline{\rule{80mm}{0.1pt}}
\vspace{2mm}

\begin{multicols}{2}

\end{multicols}

\clearpage

\end{document}